\documentclass[aps,pra,reprint,superscriptaddress, nofootinbib]{revtex4-2}

\usepackage[T1]{fontenc}
\usepackage{amsmath,amssymb,amsfonts,bm}
\usepackage{graphicx}
\usepackage{physics}
\usepackage{hyperref}
\usepackage{xcolor}
\usepackage{booktabs}
\usepackage{multirow}
\usepackage{subfigure}
\usepackage{siunitx}

\usepackage{float}
\makeatletter
\def\fnum@table{\textbf{\tablename~\thetable}}
\makeatother
\usepackage{tikz}
\usetikzlibrary{arrows.meta,decorations.pathmorphing,topaths}

\newcommand{\Gone}{\Gamma_1}
\newcommand{\GoneTLS}{\Gamma_1^{\mathrm{TLS}}}
\newcommand{\GoneBG}{\Gamma_1^{\mathrm{bg}}}
\newcommand{\tone}{T_1}
\newcommand{\ABAA}{\textsc{abaa}}
\newcommand{\DRF}{\textsc{1d-drf}}
\newcommand{\TL}{\textsc{2d-cv}}
\newcommand{\wq}{\omega_q}
\newcommand{\wTLS}{\omega_{\mathrm{TLS}}}

\makeatletter
\renewcommand{\fnum@figure}{\textbf{Figure \thefigure}}
\makeatother

\begin{document}

\title{Automating detection of Two-Level Systems in Superconducting Qubits}

\author{Rama Khalil} 
\email{rama.khalil.990@gmail.com }
\affiliation{Nordita, KTH Royal Institute of Technology and Stockholm University, Hannes Alfv\'ens v\"ag 12, SE-106 91 Stockholm, Sweden}

\author{Cameron J.\ Kopas}
\affiliation{Rigetti Computing, Berkeley, CA 94710, USA}

\author{Avinash Pathapati}
\affiliation{Nordita, KTH Royal Institute of Technology and Stockholm University, Hannes Alfv\'ens v\"ag 12, SE-106 91 Stockholm, Sweden}
\affiliation{Department of Physics, Institute for Materials Science, University of Connecticut, Storrs, Connecticut 06269, USA}

\author{Rory Cochrane}
\affiliation{Rigetti Computing, Berkeley, CA 94710, USA}

\author{Xiqiao Wang}
\affiliation{Rigetti Computing, Berkeley, CA 94710, USA}

\author{Hossein Taghinejad}
\affiliation{Rigetti Computing, Berkeley, CA 94710, USA}

\author{Vito Iaia}
\affiliation{Lawrence Livermore National Laboratory, Livermore, CA 94550}

\author{Josh Mutus}
\affiliation{Rigetti Computing, Berkeley, CA 94710, USA}

\author{Yaniv J. Rosen}
\affiliation{Lawrence Livermore National Laboratory, Livermore, CA 94550}

\author{Jared H.\ Cole}
\affiliation{Department of Physics, School of Science, RMIT University,
  Melbourne, Victoria, Australia}

\author{David P.\ Pappas}
\affiliation{Rigetti Computing, Berkeley, CA 94710, USA}

\author{Alexander V. Balatsky}
\email{balatsky@hotmail.com }
\affiliation{Nordita, KTH Royal Institute of Technology and Stockholm University,
  Hannes Alfv\'ens v\"ag 12, SE-106 91 Stockholm, Sweden}
\affiliation{Department of Physics, Institute for Materials Science,
  University of Connecticut, Storrs, Connecticut 06269, USA}

\date{\today}

\begin{abstract}
Microscopic two-level system (TLS) defects remain a primary mechanism of decoherence and operational instability in superconducting transmon qubits, necessitating scalable and automated methods for their characterization. Here, we present and benchmark two complementary analysis pipelines for extracting TLS statistics directly from time-resolved SWAP spectroscopy. One-dimensional decay-rate fitting (\DRF{}), which detects defects via localized enhancements in the qubit relaxation rate, and a deterministic, non-parametric computer-vision framework (\TL{}) that achieves two-dimensional spectral localization by exploiting the temporal persistence of coherent population suppression. We deploy both methods on SWAP spectroscopy measurements from 52 flux-tunable transmon qubits on Rigetti processors with and without moderate ($\sim 10\%$) post-fabrication frequency trimming via Alternating-Bias Assisted Annealing (\ABAA{}). We show that both pipelines converge on a consistent global characterization of the defect landscape while exhibiting complementary sensitivity across distinct coupling regimes. Crucially, both methods independently reveal a count--loss decoupling under moderate annealing: While the total detectable TLS defect density remains statistically unchanged, the frequency-integrated TLS-induced relaxation loss over the measured tuning span decreases by approximately a factor of two, demonstrating selective suppression of the most strongly dissipative defect channels. These results establish an automated, non-parametric analysis framework for high-throughput hardware diagnostics and provide a statistical baseline for post-fabrication defect engineering in large-scale superconducting quantum processors.
\end{abstract}
\maketitle

\section{\label{sec:intro}Introduction}

The realization of fault-tolerant quantum computation places stringent demands on qubit coherence. Superconducting transmon architectures, in particular, remain fundamentally limited by material-borne microscopic defects~\cite{devoret2013superconducting, kjaergaard2020stateofplay, krantz2019quantum}. Among these loss channels, microscopic two-level system (TLS) defects residing within amorphous dielectric interfaces, surface oxides, and Josephson junction barriers remain a primary bottleneck to the coherence and scalability of superconducting transmon processors~\cite{muller2019tls, phillips1987two, anderson1972anomalous, faoro2008microscopic}. When tuned into near-resonance with a transmon, these defects couple via electric dipole interactions, triggering coherent energy exchange, localized relaxation hot-spots, temporal $\tone$ fluctuations, and characteristic chevron-like avoided crossings in time-resolved spectroscopy~\cite{simmonds2004coherent, martinis2005dielectric, schlor2019correlating, klimov2018fluctuations, ColaoZanuz2025}. Systematically identifying and characterizing these defect environments across multi-qubit architectures is critical for continuous hardware calibration, defect mitigation, and processor optimization.

While time-resolved SWAP spectroscopy~\cite{lisenfeld2010measuring, Bejaninetal2021} directly maps the excited qubit population $P(|1\rangle, t, \wq)$ across interaction time and frequency, extracting statistically reliable defect populations from large-scale hardware remains an open computational challenge. Existing characterization routines typically reduce these rich two-dimensional measurement maps into isolated one-dimensional decay-rate slices $\Gone(\wq)$, discarding the spatial coherence and temporal persistence inherent in coherent avoided level crossings~\cite{Bejaninetal2021, ColaoZanuz2025, klimov2018fluctuations}. Although recent efforts have explored machine-learning regression on spectroscopic maps~\cite{mansikkamaki2024twotone, pathapati2025accelerated, KhalilThesis}, extracting defect parameters from raw, noisy experimental maps without restrictive analytical fitting or heavy supervised training priors requires a more robust, non-parametric approach.

To address this challenge, we introduce a deterministic, non-parametric two-dimensional computer-vision framework (\TL{}) designed to automate TLS defect localization directly on native SWAP spectroscopy maps. Rather than compressing the spectroscopy canvas into uncoupled 1D decay profiles,  \TL{} leverages anisotropic Gaussian scale-space filtering~\cite{witkin1983scale, lindeberg1994scale} and positive-bounded residual darkness mapping to decouple localized resonant loss from instrumentation baseline drift and flux-dependent readout envelopes. By projecting an empirical global activity mask along the interaction timeline~\cite{otsu1979threshold}, the pipeline exploits the temporal persistence of coherent state exchange, effectively integrating weak defect signatures over time while suppressing uncorrelated measurement noise. 
This establishes a deterministic, non-parametric computer-vision framework that preserves the native spatiotemporal topology of the multi-qubit measurement dataset.

We benchmark this new \TL{} framework against conventional one-dimensional decay-rate fitting (\DRF{}) and establish strong quantitative consistency between the two independent approaches, demonstrating robust per-qubit count agreement and near-unity correlation ($r=0.88$) in physical loss strengths across six orders of magnitude. Having verified the reliability of the computer-vision approach, we deploy the dual-pipeline toolchain to resolve a fundamental question in superconducting qubit defect engineering, whether microscopic TLS populations can be deterministically modified post-fabrication.

Historically, microscopic TLS defects have been regarded as static, immutable manifestations of amorphous material disorder that could not be actively reconfigured post-fabrication~\cite{phillips1987two, anderson1972anomalous, mcrae2020materials}. While thermal cycling to room temperature can stochastically reshuffle individual defect frequencies, large-scale lateral studies across multi-qubit processors show that the global defect density remains strictly conserved~\cite{ColaoZanuz2025}. The introduction of Alternating-Bias Assisted Annealing (\ABAA{})~\cite{Pappas2024ABAA} has challenged this view by demonstrating that driven atomic reorganization in $\mathrm{Al}_x\mathrm{O}_y$ barriers enables precision frequency tuning~\cite{Wang2024ABAAtuning, Tyner2025ABAAsim} alongside macroscopic loss reduction. However, a large-scale lateral study investigating the microscopic response of defect populations across a substantial multi-qubit processor has remained unavailable to date.

Equipped with our automated localization pipeline, we perform the first large-scale statistical study of TLS landscapes across 52 flux-tunable transmon qubits, comparing an as-fabricated control ensemble ($n=28$, untrimmed) against an ensemble that underwent moderate ($\sim 10\%$) frequency trimming via \ABAA{} ($n=24$, trimmed). Both diagnostic pipelines independently reveal a count--loss decoupling, that is moderate \ABAA{} treatment leaves the total detectable TLS defect count statistically unchanged while halving the span-integrated dielectric loss. By ruling out fabricated junction geometry as a confound, we prove that this loss reduction is driven by annealing-induced weakening of the most strongly coupled defect channels. Together, these results demonstrate 2D computer vision as an automated, scalable asset for superconducting qubit diagnostics and provide empirical proof of post-fabrication defect engineering across multi-qubit processors.

\section{\label{sec:theory}Transmon--TLS System}

\subsection{Microscopic Description}

Two-level system defects emerge as localized microscopic degrees of freedom in amorphous dielectric interfaces, oxide layers, and structural
disorder~\cite{phillips1987two,faoro2008microscopic,tyner2025phonon}. The
composite transmon--TLS Hamiltonian is~\cite{muller2009relaxation}
\begin{equation}
\hat{H}
=
-\tfrac{1}{2}\epsilon_q \sigma_z
-\tfrac{1}{2}\sum_n \epsilon_{f,n}\tau_{z,n}
+\tfrac{1}{2}\sigma_x\sum_n v_{\perp,n}\tau_{x,n}
+\hat{H}_{\mathrm{bath}},
\label{eq:full_hamiltonian}
\end{equation}
where $\epsilon_q$ is the qubit transition energy, $\epsilon_{f,n}$ is the
energy splitting of the $n$-th TLS, and $v_{\perp,n}$ is the transverse
qubit--TLS coupling. Here, $\sigma_{x,z}$ and $\tau_{x,z,n}$ are Pauli
operators acting on the qubit and $n$-th TLS subspaces, respectively. In the
near-resonant regime, $\epsilon_q \approx \epsilon_{f,n}$, the transverse
coupling enables coherent excitation exchange between the qubit and TLS,
giving rise to the avoided-crossing and chevron-like signatures observed in
time-resolved spectroscopy.
\begin{align}
\hat{H}_{\mathrm{bath}}
=
&\frac{1}{2}
\left(
\beta_{f,\parallel}\tau_z X_{f,\parallel}
+
\beta_{f,\perp}\tau_x X_{f,\perp}
\right)
\nonumber\\
&+
\frac{1}{2}
\left(
\beta_{q,\parallel}\sigma_z X_{q,\parallel}
+
\beta_{q,\perp}\sigma_x X_{q,\perp}
\right),
\label{eq:bath_hamiltonian}
\end{align}
where $X_{q/f,\parallel}$ and $X_{q/f,\perp}$ denote stochastic environmental bath operators driving energy relaxation ($T_1$) and pure dephasing ($T_\phi$) in the qubit and defect subspaces, while $\beta$ parameters define the respective coupling strengths to the thermal bath. 

In the near-resonant regime ($\epsilon_q \approx \epsilon_{f,n}$), the transverse coupling $v_{\perp,n}$ drives coherent excitation exchange between the transmon and localized defects. When embedded within a dissipative environment, this interaction creates localized energy relaxation hot-spots and characteristic chevron interference patterns in time-domain spectroscopy~\cite{muller2009relaxation, breuer2002theory}. An analytical derivation connecting this microscopic Hamiltonian to the open-system Lindblad master equation and effective Markovian relaxation rates is provided in Appendix~\ref{app:theory}.

\subsection{SWAP Spectroscopy}

In flux-tunable transmon architectures, the qubit transition frequency is
controlled via external magnetic flux biasing of the Josephson junction
loop~\cite{koch2007charge}. Tuning $\wq$ across a finite spectral window
brings the qubit into resonance with static environmental defects satisfying
$\wq \approx \wTLS$, triggering coherent excitation exchange~\cite{muller2009relaxation,
Bejaninetal2021}. The operational cycle proceeds in four stages~\cite{Bejaninetal2021}:

\begin{enumerate}
\item The transmon is initialized in $|10\ldots0\rangle$ via a resonant
      $\pi$-pulse. TLS defects remain unexcited.
\item A calibrated flux pulse of amplitude $A$ and duration $t$ tunes the
      qubit across the spectral window.
\item During the interaction interval, coherent excitation exchange occurs
      between the qubit and nearby resonant TLS.
\item The qubit excited-state population is measured dispersively, yielding
      the two-dimensional map $Z(\wq, t)$.
\end{enumerate}

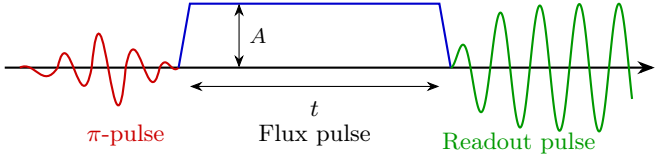
\begin{figure}[tb]
\centering
\begin{tikzpicture}[scale=1.0]
% Base timeline axis
\draw[->,>=Stealth,thick] (0,0) -- (8.6,0);

% 1. Pi-pulse (Red: Gaussian-enveloped microwave burst)
\draw[red!80!black,thick,smooth]
  (0.2,0) .. controls (0.4,0.1) and (0.5,-0.3) .. (0.7,0)
          .. controls (0.85,0.7) and (1.0,-0.8) .. (1.15,0)
          .. controls (1.3,1.6) and (1.45,-1.7) .. (1.6,0)
          .. controls (1.75,0.8) and (1.9,-0.7) .. (2.05,0)
          .. controls (2.15,0.2) and (2.2,-0.1) .. (2.3,0);
\node[red!80!black,font=\small] at (1.6,-0.9) {$\pi$-pulse};

% 2. Flux pulse (Blue: DC flat-top step)
\draw[blue!80!black,thick] (2.3,0) -- (2.45,0.85) -- (5.75,0.85) -- (5.9,0);
\node[font=\small] at (4.1,-0.9) {Flux pulse};
\draw[<->,>=Stealth] (3.1,0) -- (3.1,0.85);
\node[font=\footnotesize,fill=white,inner sep=1pt] at (3.35,0.42) {$A$};
\draw[<->,>=Stealth] (2.45,-0.25) -- (5.75,-0.25);
\node[font=\small] at (4.1,-0.55) {$t$};

% 3. Readout pulse (Green: Higher frequency microwave burst with ramp-up envelope)
\draw[green!60!black,thick,domain=5.9:8.3,samples=260,smooth] 
  plot (\x, { 0.85 * (1 - exp(-3.0*(\x-5.9))) * sin(12*(\x-5.9)*180/3.14159) });
\node[green!60!black,font=\small] at (6.8,-1.0) {Readout pulse};
\end{tikzpicture}
\caption{\textbf{SWAP spectroscopy pulse sequence.} An initial $\pi$-pulse (red) excites the qubit, a flux pulse (blue) of amplitude $A$ and duration $t$ tunes $\omega_q$ into near-resonance with candidate TLS, and a dispersive readout pulse (green) measures the qubit excited-state population.}
\label{fig:swap_sequence}
\end{figure}

\section{\label{sec:methods}Automated TLS Localization Frameworks}

Extracting reliable microscopic defect parameters from broad-band spectroscopy requires decoupling localized resonant interactions from hardware-level calibration drifts and instrumentation noise. Here, we present two complementary, automated diagnostic pipelines operating on time-resolved SWAP spectroscopy. Conventional one-dimensional decay-rate fitting (\DRF{}) and a deterministic, non-parametric two-dimensional computer-vision framework (\TL{}). While \DRF{} reduces the spectroscopy canvas to an array of independent relaxation traces, \TL{} operates natively on the full two-dimensional measurement matrix, exploiting spatiotemporal coherence to achieve spectral localization without iterative non-linear optimization. Below, we formalize the mathematical framework of both pipelines and establish a unified physical loss metric for cross-method validation.

\subsection{\label{sec:drf}1D Decay-Rate Fitting (\DRF{})}

The \DRF{} pipeline builds on the established $T_1$-based approach widely used to identify TLS resonances in superconducting qubits, in which the frequency dependence of the qubit energy-relaxation rate is used to reveal enhanced dissipation from resonant TLS coupling~\cite{Lisenfeld2019,Lisenfeld2023}. For each discrete probe frequency $f_i$, the time-resolved excited-state population $P(|1\rangle,t)$ is modeled as an unconstrained single-exponential relaxation process:
\begin{equation}
P(t) = A e^{-t/\tau_i},
\label{eq:exp_decay}
\end{equation}
where $A$ and $\tau_i$ are fitted parameters bounded by $A \in [0.1, 10]$ and $\tau \in [0, \SI{1}{\second}]$, initialized at $\tau_0 = \SI{40}{\micro\second}$. The effective frequency-dependent energy relaxation rate is subsequently defined as:
\begin{equation}
\Gone(f_i) = \frac{1}{\tau_i}.
\end{equation}

Resonant TLS candidates manifest as sharp, localized enhancements in $\Gone(f_i)$ relative to the background transmon loss profile. To isolate these excursions, a smooth background baseline is evaluated from the frequency-ordered $\{\Gone(f_i)\}$ spectrum via a rolling 15th-percentile filter over a window of $W = 50$ frequency channels. The local statistical noise floor is estimated using the rolling Median Absolute Deviation (MAD):
\begin{equation}
\sigma_i = 1.4826 \, \widetilde{\mathrm{med}}_{50} \!\left( \left| \Gone - \widetilde{\mathrm{med}}_{50}(\Gone) \right| \right)_i,
\label{eq:mad}
\end{equation}
where $\widetilde{\mathrm{med}}_k$ represents a centered rolling median of length $k$. A statistical significance boundary is defined as:
\begin{equation}
\GoneBG(f_i) = \mathrm{baseline}_i + 1.5\,\sigma_i,
\end{equation}
and the localized excess relaxation attributed to defect dissipation is isolated by positive rectification:
\begin{equation}
\GoneTLS(f_i) = \max\!\left[ \Gone(f_i) - \GoneBG(f_i), \, 0 \right].
\end{equation}
Contiguous frequency channels satisfying $\GoneTLS(f_i) > 0$ define the set of \DRF{} candidate detections. A complete schematic of the 1D fitting workflow is provided in Appendix~\ref{app:1d_pipeline}.

\subsection{\label{sec:tl}Computer-Vision-Assisted 2D Localization (\TL{})}

While 1D reduction provides a computationally tractable representation, it treats adjacent frequency channels independently and discards the coherent multi-pixel topology of avoided crossings. To overcome this limitation, we introduce \TL{}, a deterministic computer-vision framework that operates directly on the native two-dimensional population matrix $Z(t,f) \in \mathbb{R}^{N_t \times N_f}$. By leveraging multi-scale space theory~\cite{witkin1983scale, lindeberg1994scale}, \TL{} isolates coherent defect signatures via temporal persistence across the interaction duration.

\subsubsection{Anisotropic Scale-Space Background Estimation}
Raw experimental spectroscopy maps are subject to macroscopic background gradients, flux-crosstalk offsets, and slow base-temperature variations. To decouple localized avoided crossings from these non-uniform envelopes without parametric modeling, we compute a scale-space background model $B(t,f)$ using anisotropic two-dimensional Gaussian convolution:
\begin{equation}
B(t,f) = \mathcal{G}_{\sigma_t,\sigma_f} \! \left[ Z(t,f) \right],
\label{eq:gaussian_bg}
\end{equation}
where the smoothing kernel $\mathcal{G}_{\sigma_t,\sigma_f}$ is governed by asymmetric standard deviations $(\sigma_t, \sigma_f) = (80, 10)$ pixels along the temporal and spectral coordinates, respectively.

The choice of an \emph{anisotropic} kernel reflects the geometric aspect ratio of coherent avoided level crossings in time-resolved spectroscopy~\cite{lindeberg1994scale}. A coherent transmon--TLS swap chevron is spectrally narrow ($\sim \text{MHz}$, spanning only a few frequency channels) but extends over macroscopic interaction times ($\sim \SI{100}{\micro\second}$). Setting $\sigma_t = 80$ pixels ensures that the filter smooths the slow, global transmon decay envelope without smearing into sharp, vertically localized defect signatures. Concurrently, a narrow spectral bandwidth ($\sigma_f = 10$ pixels) prevents spatial over-smoothing between adjacent flux bias points.

\subsubsection{Residual Darkness Mapping}
With the baseline envelope $B(t,f)$ established, anomalous population loss is isolated via positive-bounded rectilinear deviation:
\begin{equation}
D(t,f) = \max \!\left[ B(t,f) - Z(t,f), \, 0 \right].
\label{eq:darkness}
\end{equation}
Equation~(\ref{eq:darkness}) acts as a one-sided geometric rectifier, mapping states with accelerated population depletion onto positive darkness amplitudes $D(t,f) > 0$. Positive-going instrumentation noise and readout overshoot ($Z(t,f) > B(t,f)$) are mapped identically to zero, establishing scale invariance against device-level gain shifts and baseline contrast variations.

\subsubsection{Global Activity Thresholding and Temporal Persistence}
To separate genuine defect features from baseline background noise, an empirical percentile threshold $\theta$ is evaluated over the global distribution of $D(t,f)$:
\begin{equation}
\theta = P_{85}(D).
\end{equation}
Enforcing a \emph{global} rather than a channel-resolved percentile threshold is essential to preserve spatial contrast; a per-column threshold would force an identical fraction of active pixels across every frequency slice, artificially generating false-positive detections in flat, defect-free baseline regions.

Because physical TLS defects are spectrally stationary over the duration of the measurement sequence, we exploit temporal coherence by projecting the thresholded binary activity mask along the interaction axis to define a normalized column-occupancy metric $\rho(f)$:
\begin{equation}
\rho(f) = \frac{1}{N_t} \sum_{t=1}^{N_t} \mathbf{1} \!\left[ D(t,f) > \theta \right],
\label{eq:col_occ}
\end{equation}
where $\mathbf{1}[\cdot]$ is the indicator function and $N_t$ is the total number of time steps. The metric $\rho(f) \in [0,1]$ quantifies the fraction of the measurement timeline during which a given frequency channel exhibits statistically significant population suppression. By integrating signal contributions vertically over time, $\rho(f)$ acts as a matched temporal accumulator, amplifying persistent coherent chevrons while suppressing uncorrelated, single-pixel readout fluctuations.

\subsubsection{Region Merging, Localization, and Prominence Scoring}
Candidate defect intervals are identified by enforcing a minimum persistence criterion:
\begin{equation}
\rho(f) > \rho_{\mathrm{th}},
\end{equation}
with $\rho_{\mathrm{th}} = 0.14$. Contiguous frequency channels satisfying this condition are merged into discrete spectral bounding regions $R = [f_{\mathrm{left}}, f_{\mathrm{right}}]$ via standard 1D connected-component labeling~\cite{rosenfeld1966sequential}. A minimal width constraint ($W = f_{\mathrm{right}} - f_{\mathrm{left}} + 1 \ge 1$ channel) is enforced to reject isolated single-column noise spikes.

For each candidate region $R$, the representative defect frequency $f_{\mathrm{TLS}}$ is assigned to the channel of maximum temporal persistence:
\begin{equation}
f_{\mathrm{TLS}} = \arg\max_{f \in R} \rho(f).
\end{equation}
Selecting the maximum occupancy rather than the geometric centroid ensures that $f_{\mathrm{TLS}}$ accurately tracks the physical core of the avoided level crossing.

Finally, each identified region is assigned an integrated darkness score:
\begin{equation}
S = \sum_{(t,f) \in R} D(t,f),
\label{eq:score}
\end{equation}
representing the total volume of anomalous loss in scale space and providing a deterministic metric for defect prominence ranking. Single-trace energy relaxation times $T_1(f_{\mathrm{TLS}})$ are subsequently extracted at the localized resonance coordinate using Eq.~(\ref{eq:exp_decay}). A comprehensive diagnostic diagram detailing intermediate stages of the \TL{} workflow is provided in Appendix~\ref{app:2d_pipeline}.

\subsection{\label{sec:matching}Cross-Method Matching and Classification}

Because \DRF{} and \TL{} identify TLS candidates via fundamentally distinct mathematical representations---one-dimensional decay enhancements versus two-dimensional temporal persistence---evaluating their cross-method agreement requires a spectral coincidence metric.

Let $\mathcal{F}_{\mathrm{DRF}} = \{f_i^{\mathrm{DRF}}\}$ and $\mathcal{F}_{\mathrm{CV}} = \{f_j^{\mathrm{CV}}\}$ denote the candidate frequency sets extracted by \DRF{} and \TL{}, respectively. We implement a nearest-neighbor association within a spectral matching tolerance $\delta f = \SI{25}{MHz}$:
\begin{equation}
\left| f_i^{\mathrm{DRF}} - f_j^{\mathrm{CV}} \right| \le \delta f.
\label{eq:matching_tol}
\end{equation}
The matching window $\delta f = \SI{25}{MHz}$ accounts for the finite spectral discretization of the flux sweep and the spectral breadth of strongly coupled avoided crossings, where peak decay rates and maximum temporal persistence coordinates can exhibit minor sub-linewidth offsets.

Based on Eq.~(\ref{eq:matching_tol}), the extracted features are partitioned into three disjoint subsets:
\begin{enumerate}
    \item \textbf{Consensus Detections:} Matched pairs satisfying Eq.~(\ref{eq:matching_tol}), where both independent physical observables confirm the presence of a microscopic defect.
    \item \textbf{\TL{}-Exclusive Detections:} Features identified by persistent spatiotemporal depletion ($\rho > 0.14$) that do not produce a localized $1.5\sigma$ excursion in 1D decay fits, successfully capturing coherent, weakly dissipative avoided level crossings and vacuum Rabi chevrons.
    \item \textbf{\DRF{}-Exclusive Detections:} Localized decay spikes ($\GoneTLS > 1.5\sigma$) that lack continuous vertical persistence across the SWAP pulse duration, as well as over-segmented sub-peaks resulting from single-exponential fitting on broad avoided crossings.
\end{enumerate}

\subsection{\label{sec:lorentz}Common Physical Loss Strength Quantification}

While candidate localization is method-specific, comparing defect dissipation requires quantifying physical loss on an objective, unified basis. Raw output metrics---such as $\GoneTLS$ or scale-space darkness volume $S$---reflect pipeline-dependent statistics that cannot be compared directly in physical units.

To decouple \emph{feature localization} from \emph{loss quantification}, we model the excess dissipation spectrum of all identified candidates using open-system quantum relaxation theory~\cite{muller2009relaxation, Bejaninetal2021} (detailed derivation in Appendix~\ref{app:theory}). Near resonance ($\wq \approx \wTLS$), the transverse coupling of a single TLS induces an excess relaxation profile $\GoneTLS(f)$ governed by a Lorentzian lineshape:
\begin{equation}
\GoneTLS(f) \approx \frac{A}{1 + \left( \dfrac{f - f_0}{\gamma} \right)^2},
\label{eq:lorentzian}
\end{equation}
where $A$ is the peak relaxation amplitude, $f_0$ is the fitted resonance frequency, and $\gamma$ is the half-width at half-maximum (HWHM) spectral linewidth. The total frequency-integrated TLS loss strength $\mathcal{I}_{\mathrm{TLS}}$ is given by the analytical integral:
\begin{equation}
\mathcal{I}_{\mathrm{TLS}} = \int_{-\infty}^{\infty} \GoneTLS(f) \, df = \pi A \gamma \quad [\mathrm{Hz}^2].
\label{eq:lorentzian_integral}
\end{equation}
As derived in Appendix~\ref{app:theory}, the weak-coupling limit of the open-system dynamics gives $\mathcal{I}_{\mathrm{TLS}} = \frac{\pi}{2} v_\perp^2$, proving that $\mathcal{I}_{\mathrm{TLS}}$ scales strictly with the square of the transverse coupling strength $v_\perp^2$ and is fundamentally invariant to variations in intrinsic defect decoherence or dephasing linewidths.

For candidates localized by \DRF{}, Eq.~(\ref{eq:lorentzian}) is fitted directly to the localized $\GoneTLS(f)$ spectrum. For candidates localized by \TL{}, the exact same Lorentzian model is fitted to the identical \DRF{}-derived $\GoneTLS(f)$ spectrum around the candidate coordinate $f_{\mathrm{TLS}}$. Detections from both pipelines are therefore quantified under identical optimization bounds, initializations, and frequency windows.

To ensure statistical rigor and reject noise fluctuations, a standardized amplitude significance cutoff is enforced:
\begin{equation}
A \ge \sigma_{\mathrm{local}},
\label{eq:amplitude_cut}
\end{equation}
where $\sigma_{\mathrm{local}}$ is the local MAD noise floor from Eq.~(\ref{eq:mad}). Detections failing Eq.~(\ref{eq:amplitude_cut}) are classified as non-significant baseline fluctuations and excluded from downstream loss analysis. The open-source implementation of both pipelines, matching algorithms, and synthetic map generators is available in Ref.~\cite{KhalilTLSCode}.

\section{\label{sec:results}Results}

We evaluate the \DRF{} and \TL{} diagnostic pipelines on experimental
SWAP spectroscopy maps acquired from $N_{\mathrm{qubits}} = 52$
flux-tunable transmon qubits on Rigetti processor testbeds. The dataset
comprises two distinct ensembles: an untreated baseline control group
($n=28$ qubits, ``Untrimmed'') and a post-fabrication treated group
($n=24$ qubits, ``Trimmed'') that underwent moderate ($\sim10\%$)
frequency trimming using Alternating-Bias Assisted Annealing (\ABAA{}).
We first establish the physical consistency and complementarity of the two
localization methods, and then apply them to quantify the effect of this
moderate \ABAA{} treatment on the TLS population.

\subsection{\label{sec:agreement}Cross-Method Validation}

The two localization methods identify TLS candidates using fundamentally
different representations of the same SWAP spectroscopy measurement.
Before using either method to compare the two device ensembles, we
therefore first test whether detections assigned to the same physical
resonance yield consistent measures of TLS loss strength.

\begin{figure}[tb]
\centering
\includegraphics[width=\linewidth]{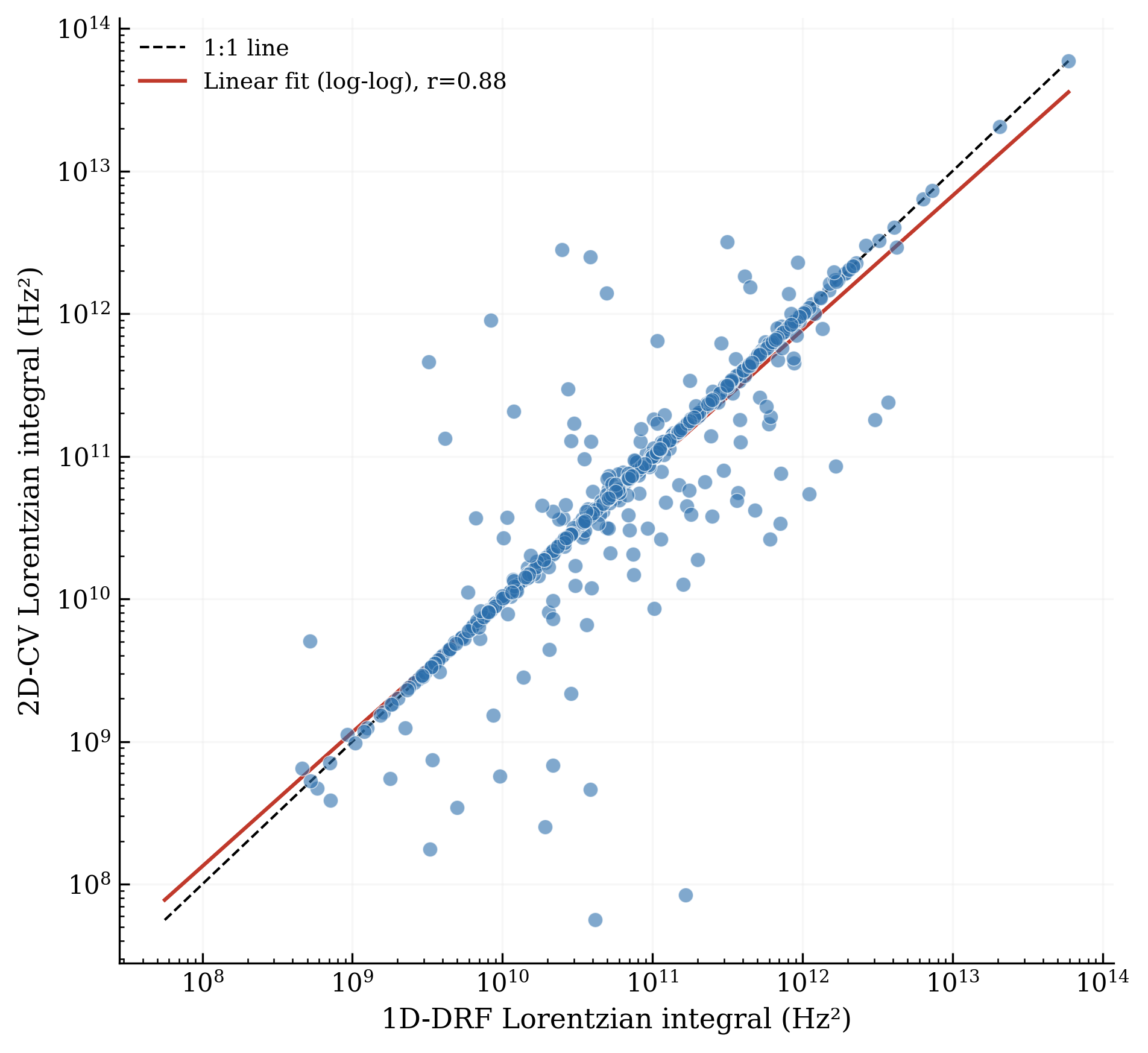}
\caption{\textbf{Matched-peak physical loss agreement across six decades.} Comparison of frequency-integrated Lorentzian loss strengths $\mathcal{I}_{\mathrm{TLS}} = \pi A \gamma$ ($\mathrm{Hz}^2$) for $n = 597$ consensus TLS pairs matched within a frequency tolerance $\delta f = \SI{25}{MHz}$. Both pipelines yield significant Lorentzian fits. Dashed line: 1:1 agreement line. Pearson correlation $r = 0.88$ on log-transformed values.}
\label{fig:lorentz_scatter}
\end{figure}

Using the nearest-neighbor matching criterion
$\delta f=\SI{25}{MHz}$ defined in Sec.~\ref{sec:matching}, we identify
$n=597$ consensus defect pairs for which statistically significant
Lorentzian fits are obtained independently by both methods. Figure~\ref{fig:lorentz_scatter}
compares the corresponding frequency-integrated loss strengths
$\mathcal{I}_{\mathrm{TLS}}=\pi A\gamma$ on a logarithmic scale.
The two methods exhibit strong agreement across approximately six orders
of magnitude, with Pearson correlation $r=0.88$ for the
log-transformed values. The consensus population lies close to the
one-to-one relation without an evident systematic offset between the two
methods.

\subsection{\label{sec:count}Invariance of Total TLS Defect Count}

Having established consistency between the two localization methods, we next apply them independently to the untrimmed and trimmed device
ensembles to determine whether moderate \ABAA{} treatment changes the
number of detectable TLS defects.

\begin{figure}[tb]
\centering
\includegraphics[width=\linewidth]{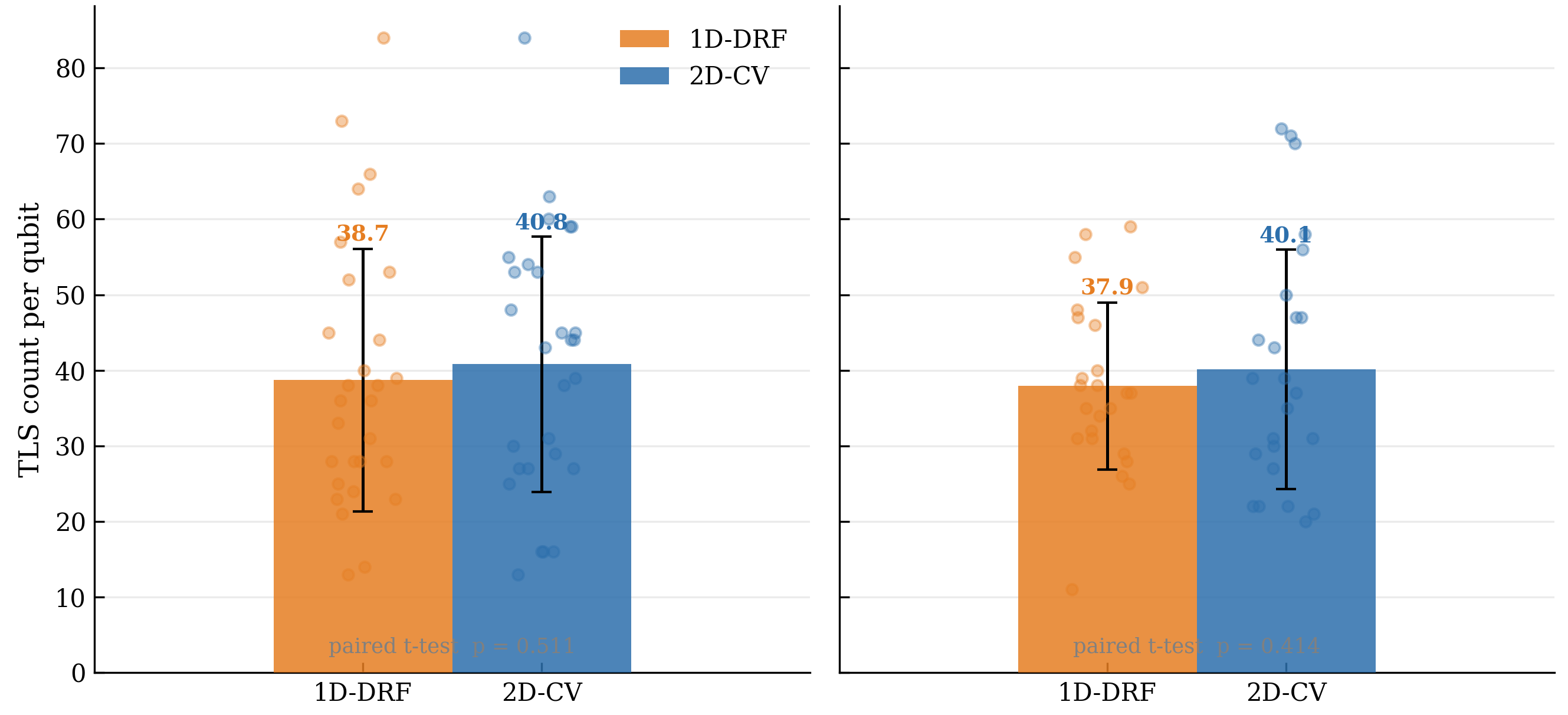}
\caption{\textbf{Mean TLS defect count per qubit (Trimmed and Untrimmed)} Comparison of detected defect counts across untreated (left) and treated (right) qubit groups evaluated independently by \DRF{} (orange) and \TL{} (blue). Columns represent group means $\pm$ standard deviation; individual dots depict per-qubit counts. Two-sample $t$-tests confirm no statistically significant change in defect density ($p > 0.8$).}
\label{fig:count}
\end{figure}

Figure~\ref{fig:count} compares the mean per-qubit TLS count across the untrimmed and trimmed device ensembles. Neither diagnostic method detects a statistically significant change in total defect count following moderate treatment. For \DRF{}, the mean defect count shifts from $38.7 \pm 17.7$ (Untrimmed) to $37.9 \pm 11.3$ (trimmed). Similarly, \TL{} yields a mean count of $40.8 \pm 17.2$ (untrimmed) versus $40.1 \pm 16.2$ (trimmed). Two-sample $t$-tests confirm that these minor shifts are statistically insignificant ($p > 0.8$, Table~\ref{tab:params}). 

Crucially, individual per-qubit defect counts span nearly an order of magnitude across the chip ($\sim 10$ to $85$ defects/qubit). This demonstrates that non-ergodic, device-to-device spatial heterogeneity dominates the total defect count variance, remaining unaffected by moderate $10\%$ frequency-trimming annealing cycles.

\subsection{\label{sec:loss}Selective Suppression of High-Dissipation TLS Candidates}

While the total number of detectable TLS defects remains invariant under moderate ($\sim10\%$) \ABAA{} treatment, evaluating the \emph{spectral loss strength} reveals an interesting physical phenomenon, that is a clear count/loss decoupling.

\begin{figure*}[tb]
\centering
\includegraphics[width=\textwidth]{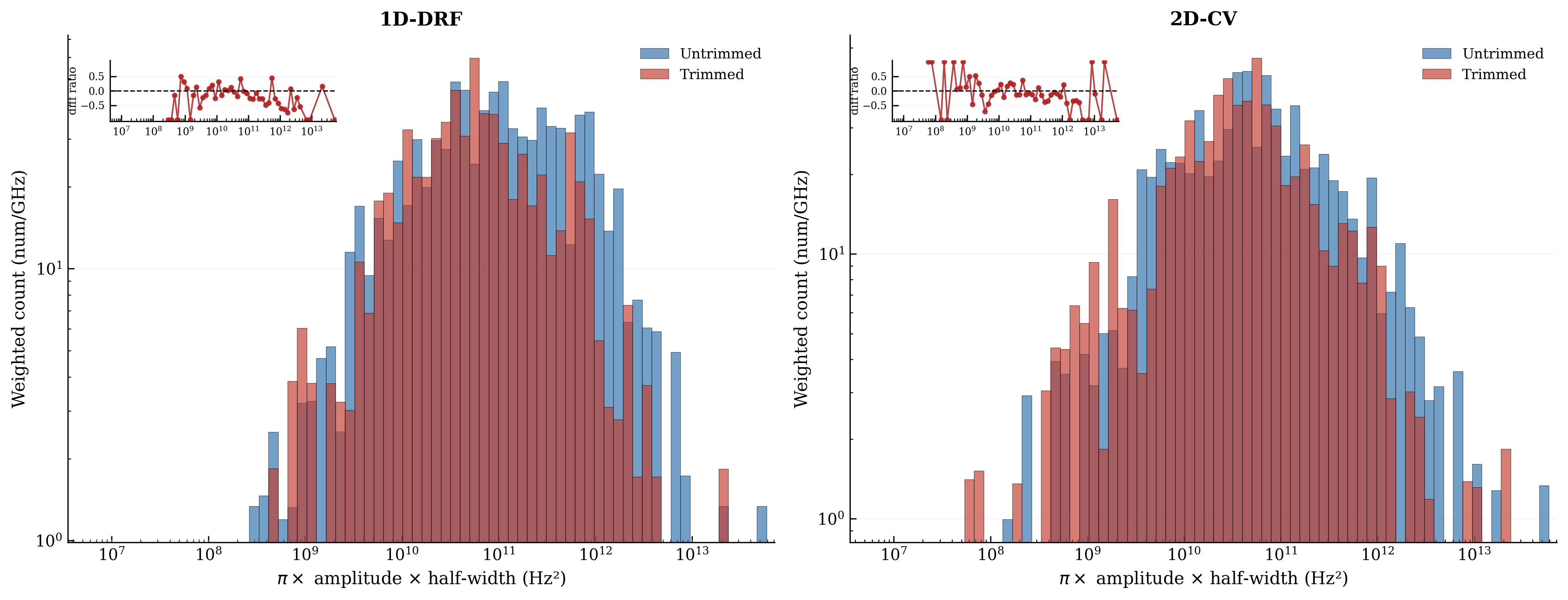}
\caption{\textbf{Lorentzian loss-strength distributions and treatment differential.} 
Main panels: Frequency-integrated loss strengths $\mathcal{I}_{\mathrm{TLS}} = \pi A \gamma$ ($\mathrm{Hz}^2$) for \DRF{} (left) and \TL{} (right), evaluated across as-fabricated baseline devices (blue, $n=28$) and \ABAA{}-trimmed devices (red, $n=24$). Only statistically significant defect fits ($A \ge \sigma_{\mathrm{local}}$) are included. 
Insets: Per-bin treatment differential ratio $(\mathcal{P}_{\mathrm{trim}} - \mathcal{P}_{\mathrm{untrim}}) / (\mathcal{P}_{\mathrm{trim}} + \mathcal{P}_{\mathrm{untrim}})$. Across both independent diagnostic pipelines, the differential ratio fluctuates near zero throughout the moderate-coupling bulk and turns systematically negative in the high-dissipation tail ($\mathcal{I}_{\mathrm{TLS}} > 10^{11}\ \mathrm{Hz}^2$), demonstrating that moderate ($\sim10\%$)\ABAA{} trimming selectively depletes strongly coupled defect channels while leaving the bulk population conserved.}
\label{fig:lorentz_integral}
\end{figure*}

Figure~\ref{fig:lorentz_integral} plots the empirical loss-strength distributions $\mathcal{I}_{\mathrm{TLS}}$ for both pipelines, restricted to significant Lorentzian fits. In both methods, the loss spectrum for untrimmed qubits forms a continuous, unimodal log-normal-like distribution spanning from $\sim 10^7$ to $10^{13}\ \mathrm{Hz}^2$. Following moderate ($\sim10\%$) \ABAA{} treatment, the high-energy tail of the distribution ($\mathcal{I}_{\mathrm{TLS}} > 10^{11}\ \mathrm{Hz}^2$) undergoes systematic depletion, shifting the defect ensemble toward lower coupling regimes.

\begin{figure}[H]
\centering
\includegraphics[width=\linewidth]{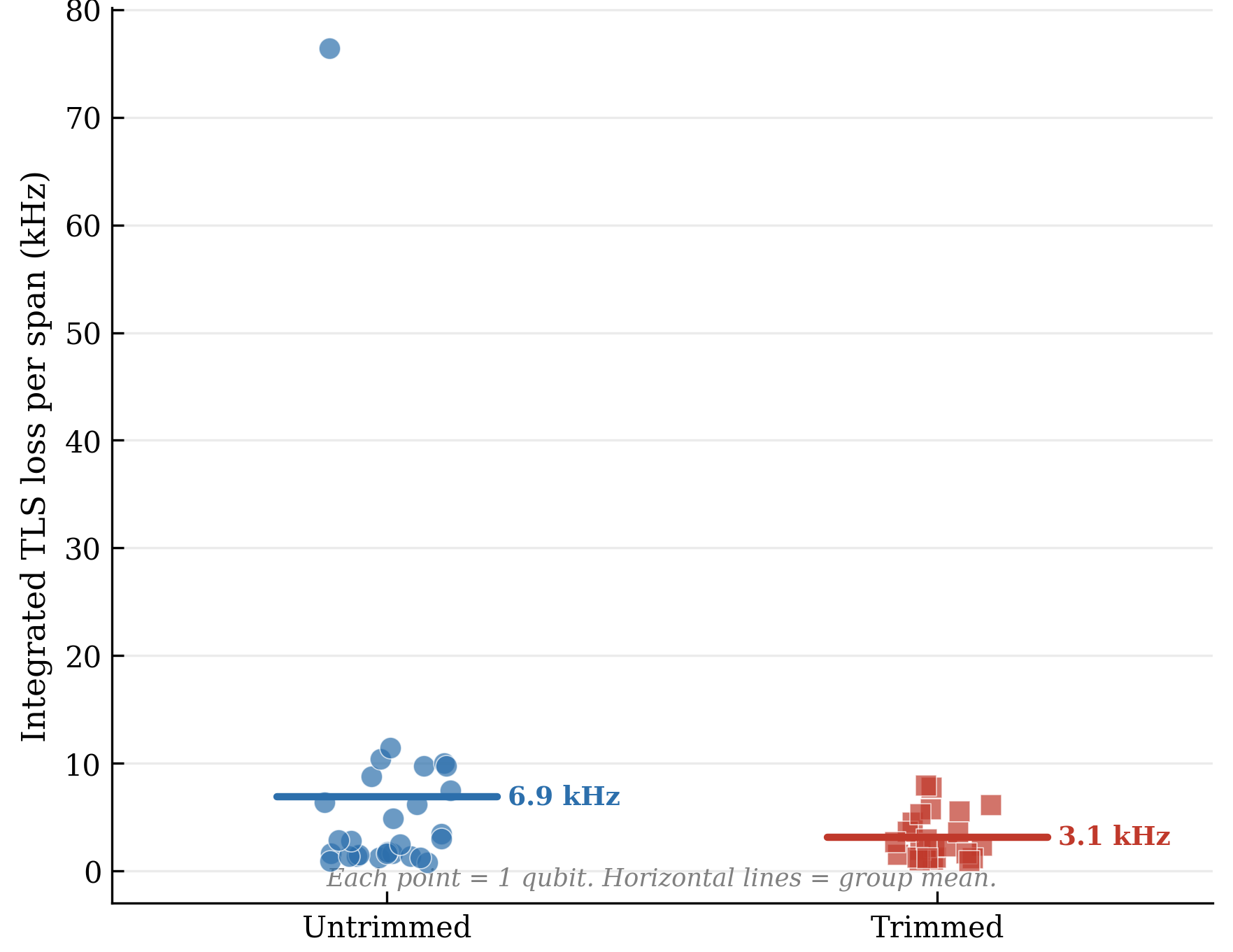}
\caption{\textbf{Span-integrated TLS dissipation per qubit.} Total integrated relaxation loss $\mathcal{L} = \sum \GoneTLS(f_i) \Delta f$ per qubit evaluated via \DRF{}. Individual dots denote single transmon devices; horizontal lines represent group ensemble means. Moderate \ABAA{} treatment reduces the mean span-integrated loss by more than $50\%$ ($6.9\ \mathrm{kHz} \to 3.1\ \mathrm{kHz}$).}
\label{fig:loss}
\end{figure}

To quantify this total loss reduction, we evaluate the span-integrated relaxation loss per qubit:
\begin{equation}
\mathcal{L} = \sum_i \GoneTLS(f_i) \, \Delta f \quad [\mathrm{kHz}].
\label{eq:span_loss}
\end{equation}
As illustrated in Fig.~\ref{fig:loss}, moderate ($\sim10\%$) \ABAA{} treatment reduces the ensemble-averaged span loss $\mathcal{L}$ by more than a factor of two, dropping from $6.9\ \mathrm{kHz}$ in untreated devices to $3.1\ \mathrm{kHz}$ in treated devices\footnote{The baseline ensemble contains a single prominent outlier at $76\ \mathrm{kHz}$ caused by an exceptionally strong, highly dissipative defect. Excluding this single outlier yields a baseline mean of $\approx 4.5\ \mathrm{kHz}$, maintaining a statistically significant loss reduction of over $30\%$}.

These results demonstrate that moderate ($\sim10\%$) \ABAA{} processing preferentially suppresses the coupling strength ($v_\perp$) of the most strongly dissipative defects rather than physically eliminating the microscopic defect states from the dielectric tunnel barrier.

Because the two device ensembles contain multiple lithographic junction
geometries, we additionally tested whether the observed loss reduction
could arise from differences in junction size rather than from the
\ABAA{} treatment itself. The two ensembles contain nine discrete SQUID
junction-area designs, with a modest imbalance in their representation.
When TLS loss strengths are resolved within individual junction-area
categories, the depletion of high-loss defects remains present for
categories containing both trimmed and untrimmed devices. A detailed geometry-stratified analysis is provided in Appendix~\ref{app:jj}. These analyses indicate that the observed loss reduction cannot be explained by the measured junction geometry alone.

\section{\label{sec:discussion}Discussion}

In this work, we bridge automated hardware diagnostics and the microscopic physics of defect engineering in superconducting quantum circuits. By executing a large-scale lateral study across 52 flux-tunable transmon qubits, we simultaneously establish the validity of a deterministic, non-parametric two-dimensional computer-vision framework (\TL{}) and provide empirical proof of defect modification after moderate ($\sim 10\%$) frequency trimming via \ABAA{}.

The central methodological advantage of \TL{} lies in its direct exploitation of the native two-dimensional measurement maps $Z(t,f)$. Conventional automated characterization routines routinely project time-resolved spectroscopy onto uncoupled one-dimensional decay-rate profiles $\Gone(\wq)$~\cite{Bejaninetal2021, ColaoZanuz2025, klimov2018fluctuations}. While computationally expedient for isolated, incoherent loss spikes, this 1D reduction fundamentally discards the temporal persistence and spatial coherence of avoided crossings. Furthermore, non-linear least-squares fitting of single-exponential decay models $P(t) = A e^{-t/\tau}$ encounters intrinsic model mismatch when forced onto oscillatory vacuum Rabi chevrons, inflating parameter variance and causing fit divergence in low-signal regimes. In contrast, \TL{} operates deterministically in scale space~\cite{witkin1983scale, lindeberg1994scale}. By applying anisotropic Gaussian convolution and empirical activity thresholding, \TL{} functions as a matched temporal accumulator that integrates coherent population suppression over the interaction timeline, enhancing defect sensitivity without requiring supervised neural network priors or restrictive analytical fitting models.

Crucially, cross-validating \TL{} against standard \DRF{} establishes both rigorous quantitative consistency and physical complementarity across distinct loss regimes. For consensus defects ($n=597$ matched pairs within $\delta f = \SI{25}{\mega\hertz}$), both independent pipelines exhibit near-unity correlation ($r=0.88$) in physical loss strengths $\mathcal{I}_{\mathrm{TLS}} = \pi A \gamma$ spanning six orders of magnitude (Fig.~\ref{fig:lorentz_scatter}). At the same time, the two methods exhibit orthogonal sensitivity profiles (Appendix~\ref{app:t1}): while consensus detections correspond to severe $T_1$ bottlenecks across the processor, \TL{}-exclusive detections successfully capture weakly dissipative fluctuators and coherent state-exchange chevrons whose lifetimes remain close to the baseline ($T_1^{\mathrm{TLS}} / T_1^{\mathrm{bg}} \approx 0.85\text{--}0.95$). Because \TL{} extracts these weak, coherent features deterministically in closed form, it provides an efficient, divergence-free screening engine for large-scale multi-qubit hardware.

While our finding that the mean detectable defect density remains conserved ($\sim 39\ \text{defects/qubit}$, $p > 0.8$, Fig.~\ref{fig:count}) corroborates the established understanding that the macroscopic density of tunneling states is an inherent feature of amorphous disorder~\cite{phillips1987two, anderson1972anomalous, muller2019tls, mcrae2020materials, ColaoZanuz2025}, evaluating spectral dissipation reveals a striking departure from a static loss picture. Specifically, our comparative analysis between as-fabricated control devices and moderately ($\sim 10\%$) trimmed devices uncovers a distinct \emph{count--loss decoupling}: despite constant defect counts, the span-integrated dielectric loss is reduced by more than a factor of two ($6.9\ \mathrm{kHz} \to 3.1\ \mathrm{kHz}$, Fig.~\ref{fig:loss}).

This selective suppression provides insight into the microscopic mechanisms governing electric-field-assisted barrier annealing~\cite{Pappas2024ABAA, Tyner2025ABAAsim}. Under alternating electrical bias at room temperature, driven ionic migration via a Cabrera-Mott mechanism reorganizes local atomic coordination within the amorphous $\mathrm{Al}_x\mathrm{O}_y$ tunnel barrier. The empirical depletion of the high-energy loss tail ($\mathcal{I}_{\mathrm{TLS}} > 10^{11}\ \mathrm{Hz}^2$, Fig.~\ref{fig:lorentz_integral}) indicates that strongly dissipative defect channels—which typically reside in regions of high local strain or elevated electric field concentration—exhibit a lower activation barrier for bias-induced structural reconfiguration. As derived in Appendix~\ref{app:theory}, because the frequency-integrated loss strength $\mathcal{I}_{\mathrm{TLS}} = \frac{\pi}{2} v_\perp^2$ scales strictly with the square of the transverse dipole coupling, this suppression represents an intrinsic physical attenuation of the microscopic interaction matrix element $v_\perp$ rather than a superficial artifact of defect linewidth broadening or thermal scrambling. Furthermore, our area-stratified robustness checks (Appendix~\ref{app:jj}) confirm that this loss reduction persists within identical lithographic SQUID geometries, ruling out physical junction dimensions as a confounding factor.

Beyond its application to barrier annealing, the \TL{} framework provides an open-source, scalable foundation for high-throughput processor characterization~\cite{KhalilTLSCode}. Because its scale-space and temporal projection operations depend on general spatiotemporal coherence rather than device-specific calibration models, the pipeline is readily extensible to broader multi-qubit architectures, continuous automated retuning routines, and wafer-scale quality assurance protocols.

\section{\label{sec:conclusions}Conclusions}

We have established and experimentally benchmarked a deterministic, non-parametric two-dimensional computer-vision framework (\TL{}) for automated TLS defect localization in superconducting quantum processors. By operating natively on two-dimensional SWAP spectroscopy maps, \TL{} leverages anisotropic scale-space filtering and temporal persistence projection to isolate coherent avoided level crossings directly, eliminating the convergence failures and parameter biases inherent in uncoupled 1D single-exponential fitting.

Deploying \TL{} alongside conventional decay-rate fitting across 52 flux-tunable transmon qubits confirms the quantitative consistency of both pipelines ($r=0.88$ over six decades in physical loss strength) while demonstrating the enhanced sensitivity of 2D computer vision to coherent, low-dissipation defect structures.

Applying this dual-pipeline architecture to compare as-fabricated control devices with moderately trimmed hardware cohorts provides the first large-scale statistical evidence of post-fabrication defect engineering: moderate ($\sim 10\%$) frequency trimming via \ABAA{} preserves the overall defect density while reducing total dielectric loss by a factor of two through the selective suppression of the most strongly coupled loss channels.

These results establish non-parametric computer vision as a scalable tool for high-throughput quantum hardware diagnostics and provide a validated empirical baseline for deterministic defect engineering in fault-tolerant superconducting architectures.

\begin{acknowledgments}
We are grateful to O. Mansikamaki and J. Sauls for useful discussions.
This work was supported by USAFOSR Grant No.~AF9550-25-1-0103 and was
performed in part under the auspices of the U.S. Department of Energy by
Lawrence Livermore National Laboratory under Contract No.~89233218CNA000001.
\end{acknowledgments}

\appendix

\section{\label{app:theory}Microscopic Hamiltonian and Open-System Loss Derivation}

This appendix provides the theoretical connection between the microscopic
transmon--TLS Hamiltonian introduced in Sec.~\ref{sec:theory} and the
Lorentzian excess-relaxation model used to quantify TLS loss in
Sec.~\ref{sec:lorentz}. We consider the weak-coupling, near-resonant regime
in which a TLS acts as an additional dissipative channel for the qubit.

\subsection{Effective Qubit--TLS Subspace}

The microscopic Hamiltonian introduced in Eq.~(\ref{eq:full_hamiltonian})
contains a transverse qubit--TLS interaction
\begin{equation}
\hat{H}_{\mathrm{int}}
=
\frac{1}{2}v_{\perp}
\sigma_x\tau_x.
\end{equation}
Near resonance, the rapidly oscillating counter-rotating terms can be
neglected within the rotating-wave approximation. The interaction then
takes the exchange form
\begin{equation}
\hat{H}_{\mathrm{int}}^{\mathrm{RWA}}
=
\frac{\hbar v_{\perp}}{2}
\left(
\sigma_+\tau_-+
\sigma_-\tau_+
\right),
\label{eq:app_exchange}
\end{equation}
where $\sigma_{\pm}$ and $\tau_{\pm}$ are the qubit and TLS ladder
operators, respectively. The corresponding single-excitation
Hamiltonian is therefore
\begin{equation}
\hat{H}_{\mathrm{sub}}
=
\frac{\hbar\omega_q}{2}\sigma_z
+
\frac{\hbar\omega_{\mathrm{TLS}}}{2}\tau_z
+
\frac{\hbar v_{\perp}}{2}
\left(
\sigma_+\tau_-+
\sigma_-\tau_+
\right),
\label{eq:app_subspace_ham}
\end{equation}
with $\omega_q$ and $\omega_{\mathrm{TLS}}$ denoting angular transition
frequencies. The resulting eigenfrequencies are
\begin{equation}
\omega_{\pm}
=
\frac{\omega_q+\omega_{\mathrm{TLS}}}{2}
\pm
\frac{1}{2}
\sqrt{
\Delta^2+v_{\perp}^2
},
\end{equation}
where
\begin{equation}
\Delta=\omega_q-\omega_{\mathrm{TLS}}.
\end{equation}
Thus, near resonance, the qubit--TLS coupling produces the avoided
crossings and coherent excitation exchange observed in time-resolved
spectroscopy~\cite{muller2009relaxation,Bejaninetal2021}.

\subsection{Open-System Dynamics}

The qubit and TLS are coupled to environmental degrees of freedom through
the bath Hamiltonian in Eq.~(\ref{eq:bath_hamiltonian}). Under the standard
Born--Markov approximation, these environmental channels can be described
by a Lindblad master equation~\cite{breuer2002theory}:
\begin{align}
\dot{\rho}
=&-\frac{i}{\hbar}
[\hat{H}_{\mathrm{sub}},\rho]
+\mathcal{D}
[\sqrt{\Gamma_{1,q}}\,\sigma_-]\rho
+\mathcal{D}
[\sqrt{\Gamma_{1,\mathrm{TLS}}}\,\tau_-]\rho
\nonumber\\
&+
\mathcal{D}
\left[
\sqrt{\frac{\Gamma_{\phi,\mathrm{TLS}}}{2}}\,
\tau_z
\right]\rho ,
\label{eq:app_lindblad}
\end{align}
where
\begin{equation}
\mathcal{D}[\hat{L}]\rho
=
\hat{L}\rho\hat{L}^{\dagger}
-\frac{1}{2}
\left\{
\hat{L}^{\dagger}\hat{L},\rho
\right\}.
\end{equation}
Here $\Gamma_{1,q}$ is the intrinsic qubit energy-relaxation rate,
$\Gamma_{1,\mathrm{TLS}}$ is the intrinsic TLS relaxation rate, and
$\Gamma_{\phi,\mathrm{TLS}}$ is its pure-dephasing rate. The corresponding
transverse decoherence rate of the TLS is
\begin{equation}
\gamma_2
=
\frac{\Gamma_{1,\mathrm{TLS}}}{2}
+
\Gamma_{\phi,\mathrm{TLS}}.
\label{eq:app_gamma2}
\end{equation}

In the weak-coupling regime, the TLS coherence can be adiabatically
eliminated from the coupled dynamics. Equivalently, second-order
perturbation theory gives the additional qubit relaxation channel produced
by the near-resonant TLS. With the coupling convention of
Eq.~(\ref{eq:app_exchange}), the resulting excess qubit relaxation rate is
\begin{equation}
\Gamma_{1,\mathrm{TLS}}(\omega_q)
=
\frac{v_{\perp}^{\,2}}{2}
\frac{\gamma_2}
{
(\omega_q-\omega_{\mathrm{TLS}})^2+\gamma_2^2
}.
\label{eq:app_lorentzian_angular}
\end{equation}
This expression describes the Lorentzian enhancement of the qubit
relaxation rate as the qubit frequency approaches the TLS resonance.

\subsection{Lorentzian Form in Frequency Units}

The experimental analysis in Sec.~\ref{sec:lorentz} is expressed in ordinary
frequency $f$ rather than angular frequency $\omega$. We therefore define
\begin{equation}
f=\frac{\omega}{2\pi},
\qquad
f_{\mathrm{TLS}}=\frac{\omega_{\mathrm{TLS}}}{2\pi},
\qquad
g=\frac{v_{\perp}}{2\pi},
\qquad
\gamma_f=\frac{\gamma_2}{2\pi}.
\end{equation}
Using these definitions, Eq.~(\ref{eq:app_lorentzian_angular}) becomes
\begin{equation}
\Gamma_{1,\mathrm{TLS}}(f)
=
\frac{g^2}{2}
\frac{\gamma_f}
{
(f-f_{\mathrm{TLS}})^2+\gamma_f^2
}.
\label{eq:app_lorentzian_frequency}
\end{equation}
This can be written in the form used in the main text,
\begin{equation}
\Gamma_{1,\mathrm{TLS}}(f)
=
\frac{A}
{
1+
\left(
\dfrac{f-f_{\mathrm{TLS}}}{\gamma_f}
\right)^2
},
\label{eq:app_lorentzian_main}
\end{equation}
with
\begin{equation}
A=\frac{g^2}{2\gamma_f}.
\label{eq:app_amplitude}
\end{equation}
Thus, the fitted amplitude and linewidth are not independent microscopic
parameters: within this simplified model their combination is determined
by the transverse qubit--TLS coupling.

\subsection{Integrated Loss Strength}

The frequency-integrated excess relaxation rate follows directly from the
Lorentzian form:
\begin{equation}
\mathcal{I}_{\mathrm{TLS}}
=
\int_{-\infty}^{\infty}
\Gamma_{1,\mathrm{TLS}}(f)\,df
=
\pi A\gamma_f .
\label{eq:app_integral}
\end{equation}
Substituting Eq.~(\ref{eq:app_amplitude}) gives
\begin{equation}
\boxed{
\mathcal{I}_{\mathrm{TLS}}
=
\frac{\pi}{2}g^2
}
\qquad [\mathrm{Hz}^2].
\label{eq:app_integral_coupling}
\end{equation}
Therefore, within the weak-coupling Lorentzian model, the integrated loss
strength is proportional to the square of the transverse qubit--TLS
coupling and does not depend explicitly on the TLS decoherence rate
$\gamma_f$. Changes in $\gamma_f$ redistribute the same integrated
spectral weight between peak amplitude and linewidth.

Equation~(\ref{eq:app_integral_coupling}) provides the physical motivation
for using $\mathcal{I}_{\mathrm{TLS}}=\pi A\gamma_f$ as the common loss
metric in Sec.~\ref{sec:lorentz}: although the two localization pipelines
use different observables to identify candidate TLS, their candidates can
subsequently be quantified through the same Lorentzian physical model.

The analytical identity above assumes integration over the full Lorentzian
line. In the experiment, the fitted Lorentzian is evaluated over the
available spectral window, so the extracted quantity should be understood
as the corresponding full-line Lorentzian estimate rather than a direct
numerical integral of the measured spectrum.

\section{\DRF{} Localization Pipeline}
\label{app:1d_pipeline}

\begin{figure*}[t]
    \centering
    \includegraphics[width=\textwidth]{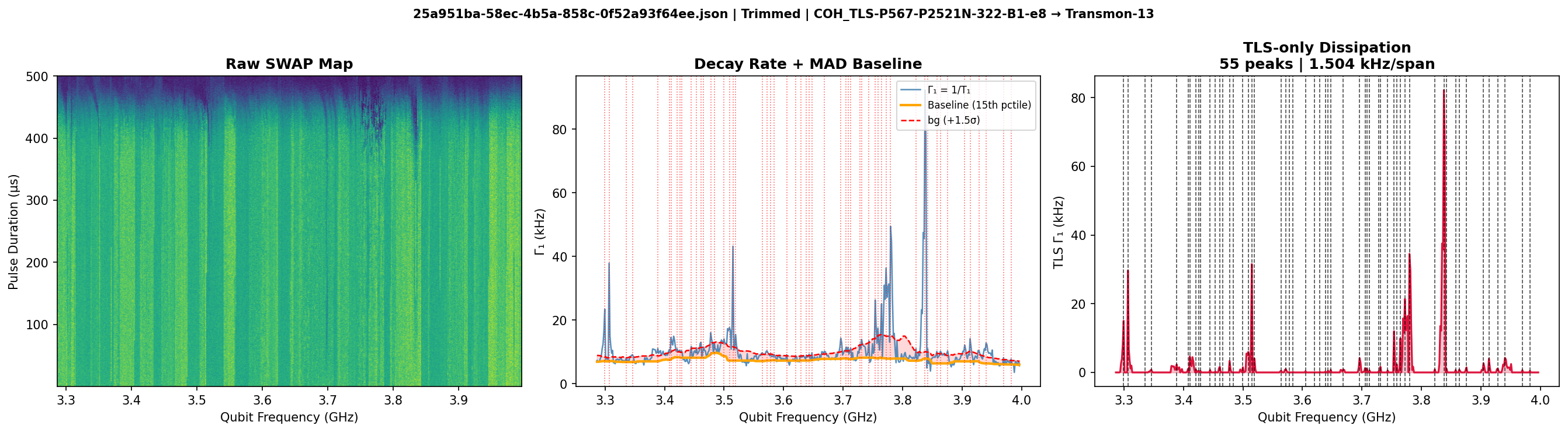}
    \caption{\textbf{Workflow of the \DRF{} localization pipeline.} 
    Left panel: Raw time-resolved SWAP spectroscopy population map $Z(t,f)$ for a representative trimmed transmon device (Transmon-13) across a \SI{700}{\mega\hertz} bandwidth. 
    Middle panel: Extracted single-exponential energy relaxation rate spectrum $\Gone(f) = 1/T_1(f)$ (blue trace), overlaid with the rolling 15th-percentile background model (orange curve) and the statistical detection floor $\GoneBG = \text{baseline} + 1.5\sigma_{\mathrm{MAD}}$ (dashed red curve). Vertical dotted lines denote channels passing thresholding. 
    Right panel: Isolated excess TLS dissipation spectrum $\GoneTLS(f) = \max[\Gone - \GoneBG, 0]$, identifying 55 candidate loss peaks yielding an integrated span loss of \SI{1.504}{\kilo\hertz}.}
    \label{fig:appendix_1d_pipeline}
\end{figure*}

The one-dimensional decay-rate fitting (\DRF{}) pipeline automates defect characterization by treating each probe frequency slice as an uncoupled temporal relaxation problem. Figure~\ref{fig:appendix_1d_pipeline} details the operational execution sequence across a representative experimental SWAP map.

For each frequency channel $f_i$, the pipeline executes a bounded non-linear least-squares fit (Levenberg-Marquardt) to the single-exponential form $P(t) = A e^{-t/\tau_i}$. When a strong, incoherent decay channel dominates, this single-exponential parameterization converges rapidly. However, two inherent operational limitations arise when applying 1D single-exponential fitting across broad-band spectroscopy:
\begin{enumerate}
    \item \textbf{Coherent Dynamics Mismatch:} Near strongly coupled avoided level crossings, transmon--TLS state hybridization produces coherent population exchange (vacuum Rabi oscillations) rather than monotonic exponential decay. Forcing a single-exponential decay model on an oscillatory time trace degrades goodness-of-fit metrics and increases parameter uncertainty.
    \item \textbf{Sub-Threshold Coherent Loss:} Weakly coupled defects produce shallow, vertically persistent population depletion that perturbs the extracted single-column rate $\Gone(f_i)$ by less than the local $1.5\sigma$ MAD noise floor. Because 1D reduction evaluates each frequency channel independently, it cannot leverage the multi-pixel spatiotemporal persistence necessary to resolve these subtle coherent features.
\end{enumerate}

\section{\TL{} Localization Pipeline}
\label{app:2d_pipeline}

\begin{figure*}[t]
    \centering
    \includegraphics[width=\textwidth]{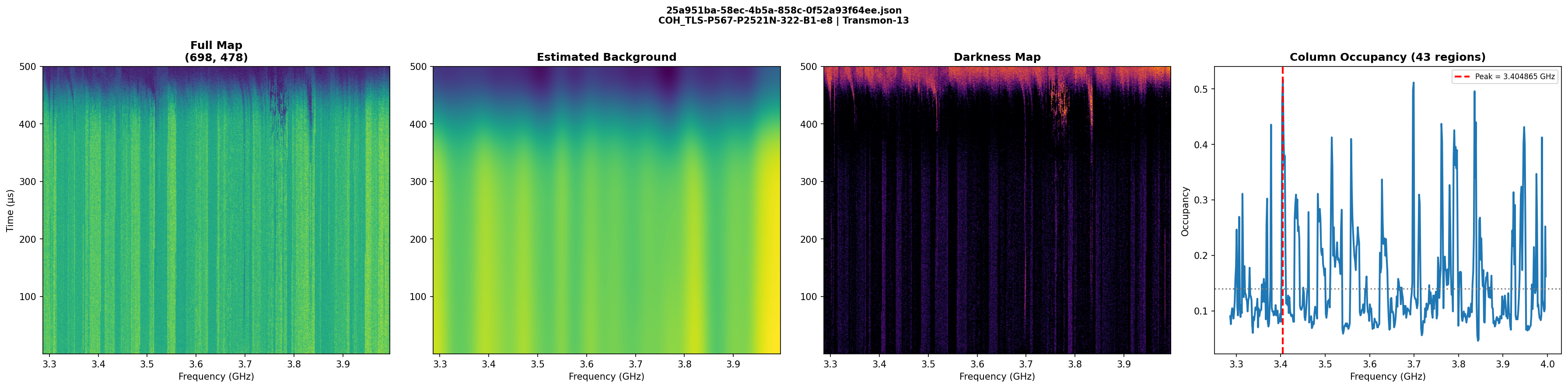}
    \caption{\textbf{Workflow of the \TL{} pipeline.} 
    From left to right: 
    (1) Native experimental SWAP population map. 
    (2) Non-parametric background baseline model $B(t,f)$ generated via anisotropic Gaussian scale-space convolution with kernel dimensions $(\sigma_t, \sigma_f) = (80, 10)$ pixels. 
    (3) Positive-rectified residual darkness map $D(t,f) = \max[B(t,f) - Z(t,f), 0]$, isolating anomalous defect-induced population suppression while mapping instrumentation readout noise to zero. 
    (4) Normalized column-occupancy projection $\rho(f)$ obtained by evaluating temporal persistence of pixels exceeding the global percentile threshold $\theta = P_{85}(D)$. The horizontal dotted line marks the candidate detection threshold $\rho_{\mathrm{th}} = 0.14$; the dashed red line marks the global maximum occupancy peak ($f = \SI{3.404865}{\giga\hertz}$, $\rho \approx 0.51$).}
    \label{fig:appendix_2d_pipeline}
\end{figure*}

\begin{figure*}[t]
    \centering
    \includegraphics[width=0.88\textwidth]{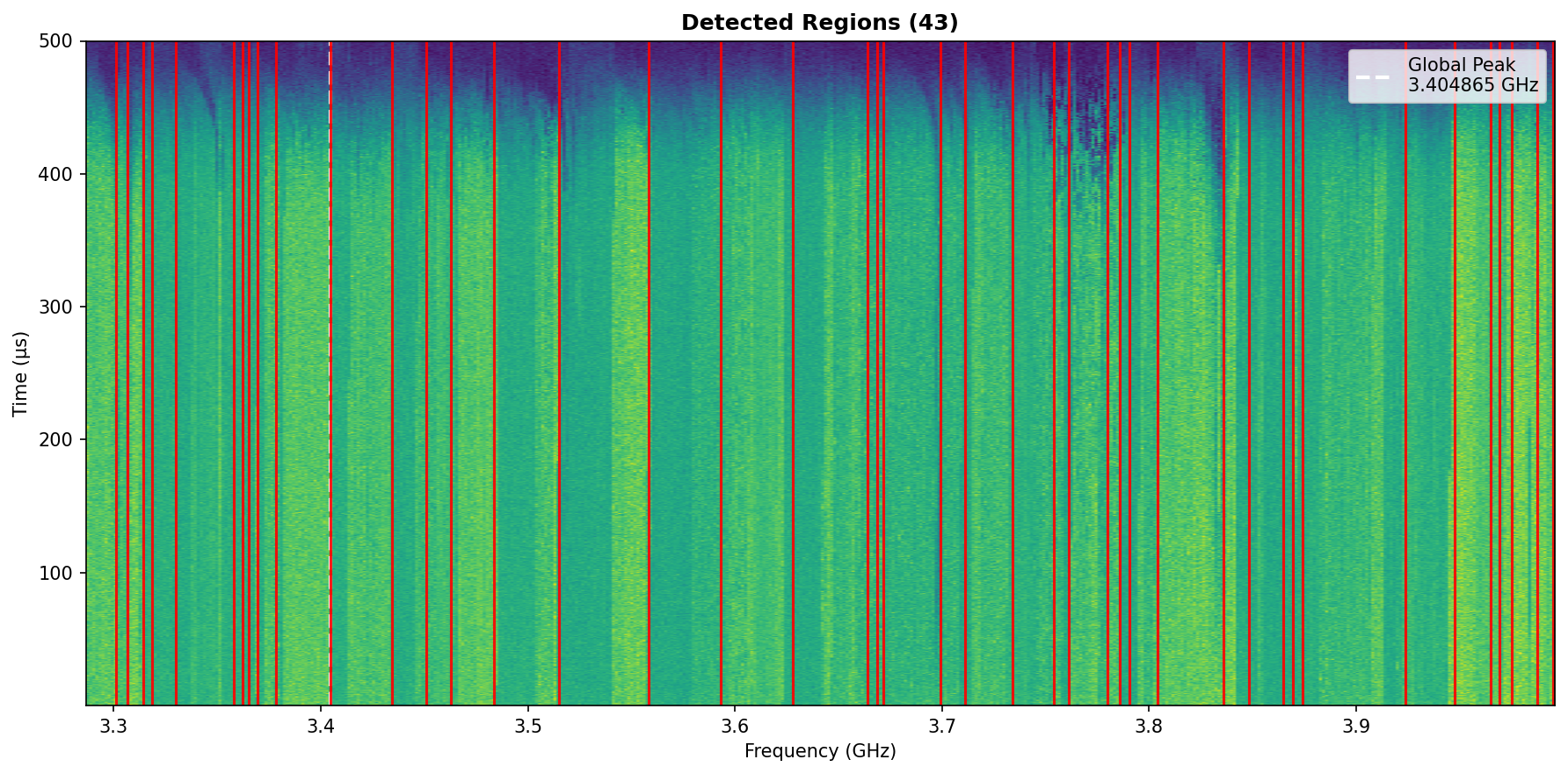}
    \caption{\textbf{Automated 2D spectral bounding and defect localization via \TL{}.} 
    Raw experimental SWAP map with all 43 automated candidate defect regions overlaid as solid red vertical lines, positioned at their maximal temporal persistence coordinates $f_{\mathrm{TLS}} = \arg\max_{f \in R} \rho(f)$. The dashed white line denotes the primary global defect at $f = \SI{3.404865}{\giga\hertz}$, corresponding to the maximum occupancy score. Spectral candidate bounds are extracted deterministically without parametric curve fitting.}
    \label{fig:tl_regions}
\end{figure*}

The two-dimensional computer-vision framework (\TL{}) addresses the limitations of 1D curve fitting by formulating defect identification as a deterministic spatiotemporal feature segmentation task. Figure~\ref{fig:appendix_2d_pipeline} illustrates the intermediate matrix transformations of the pipeline, while Figure~\ref{fig:tl_regions} demonstrates the resulting automated spectral bounding on raw data.

\section{\label{app:parameters}Analysis Parameters and Sensitivity}

To eliminate selection bias and ensure computational reproducibility, all algorithmic hyperparameters and detection thresholds were fixed prior to multi-qubit processing and held strictly constant across all 52 transmon devices. Table~\ref{tab:params} provides a comprehensive compilation of all operational hyperparameters used across both pipelines.

\begin{table}[H]
\caption{\label{tab:params}
\textbf{Fixed algorithmic hyperparameters and evaluation thresholds.}
All computational parameters were held strictly constant across all 52 transmon devices and both treatment ensembles.}
\begin{ruledtabular}
\begin{tabular}{llc}
\textbf{Parameter} & \textbf{Value} & \textbf{Pipeline} \\
\hline
\textbf{Rolling window} $W$              & 50 freq.\ pts       & \DRF{} \\
\textbf{Baseline quantile}               & 15th percentile     & \DRF{} \\
\textbf{MAD scale $k$}                   & 1.5                 & \DRF{} \\
\textbf{$\tau_0$ (initial guess)}        & \SI{40}{\micro\second} & Both \\
\textbf{$\sigma_f$ (Gaussian blur)}      & 10 pixels           & \TL{} \\
\textbf{$\sigma_t$ (Gaussian blur)}      & 80 pixels           & \TL{} \\
\textbf{Darkness percentile $\theta$}    & 85th percentile     & \TL{} \\
\textbf{Min.\ column occupancy $\rho$}   & 0.14                & \TL{} \\
\textbf{Min.\ region width}              & 1 column            & \TL{} \\
\textbf{Match tolerance $\delta f$}      & \SI{25}{MHz}        & Matching \\
\end{tabular}
\end{ruledtabular}
\end{table}

\section{\label{app:t1}Qubit Relaxation Dynamics}

To establish the physical mechanisms underlying the detection sensitivity of \DRF{} and \TL{}, we evaluate the single-trace energy relaxation times $T_1$ extracted at detected defect resonances relative to off-resonant baseline levels across all 52 transmon devices.

\begin{figure*}[tb]
    \centering
    \includegraphics[width=0.9\textwidth]{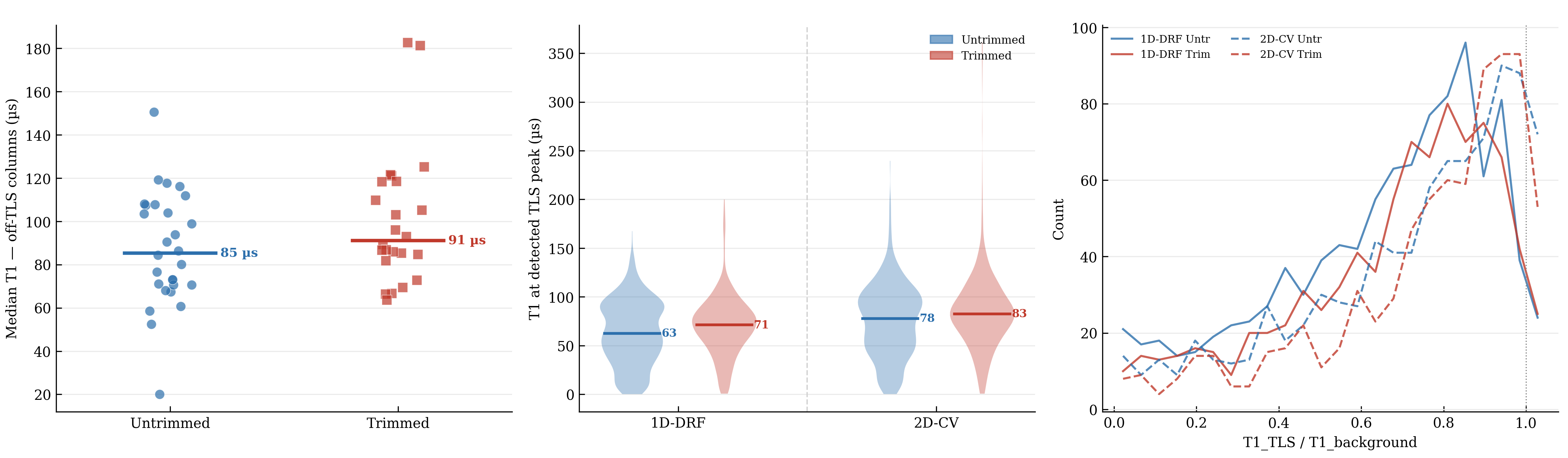}
    \caption{\textbf{Qubit relaxation lifetime characterization across baseline and defect-affected channels.} 
    Left panel: Median off-resonant background transmon lifetime per qubit ($T_1^{\mathrm{bg}}$) evaluated across unperturbed channels ($\GoneTLS = 0$) for untrimmed ($85\ \mu\mathrm{s}$, blue dots) and trimmed ($91\ \mu\mathrm{s}$, red squares) ensembles. Horizontal bars denote ensemble medians. 
    Middle panel: Empirical distributions of extracted relaxation times $T_1$ at detected defect coordinates for \DRF{} (median: $63\ \mu\mathrm{s}$ untrimmed, $71\ \mu\mathrm{s}$ trimmed) and \TL{} (median: $78\ \mu\mathrm{s}$ untrimmed, $83\ \mu\mathrm{s}$ trimmed). 
    Right panel: Defect lifetime depression ratio $T_1^{\mathrm{TLS}} / T_1^{\mathrm{bg}}$ for \DRF{} (solid lines) and \TL{} (dashed lines) across untrimmed and trimmed groups. Values near unity indicate weak, coherent loss channels.}
    \label{fig:t1_analysis}
\end{figure*}

Figure~\ref{fig:t1_analysis} (left panel) first confirms that the post-fabrication \ABAA{} process does not degrade the intrinsic transmon coherence. Evaluating the median $T_1$ across off-resonant background channels ($\GoneTLS = 0$) yields an ensemble median of $85\ \mu\mathrm{s}$ for untrimmed transmons and $91\ \mu\mathrm{s}$ for trimmed transmons, demonstrating stable baseline dielectric performance across the processor.

Figures~\ref{fig:t1_analysis} (middle panel) and \ref{fig:t1_analysis} (right panel) reveal distinct physical sensitivity profiles between the two detection pipelines. Because \DRF{} requires localized decay rates to breach a statistical threshold ($\Gone > \text{baseline} + 1.5\sigma_{\mathrm{MAD}}$), it preferentially isolates severe $T_1$ degradation events, resulting in lower median lifetimes at detected peaks ($63\ \mu\mathrm{s}$ untrimmed, $71\ \mu\mathrm{s}$ trimmed). 

In contrast, \TL{} identifies candidate regions by evaluating spatiotemporal persistence ($\rho(f) > 0.14$) across the entire interaction window. As a consequence, \TL{} successfully captures weakly dissipative fluctuators and coherent avoided crossings whose lifetimes remain close to the background baseline ($T_1^{\mathrm{TLS}} / T_1^{\mathrm{bg}} \approx 0.85\text{--}0.95$), yielding higher median peak lifetimes ($78\ \mu\mathrm{s}$ untrimmed, $83\ \mu\mathrm{s}$ trimmed). This demonstrates that \TL{} operates with enhanced sensitivity to coherent, low-dissipation defect structures that evade conventional 1D single-exponential rate thresholding.

\section{\label{app:jj}Exclusion of Junction Geometry Confounders}

Because the untrimmed and trimmed device ensembles comprise nine discrete lithographic SQUID junction designs, we evaluate whether nominal variations in junction area could account for the observed reduction in TLS dissipation. 

Independent room-temperature scanning electron microscopy and resistance measurements reveal minor dimensional offsets between the two ensembles. The mean dimensions of the two constituent junctions are $209.4 \pm 7.3\ \mathrm{nm}$ versus $213.0 \pm 5.8\ \mathrm{nm}$ for JJ1 ($p = 0.064$, two-sided Welch's $t$-test) and $109.6 \pm 3.1\ \mathrm{nm}$ versus $111.3 \pm 2.5\ \mathrm{nm}$ for JJ2 ($p = 0.033$) in the untrimmed and trimmed groups, respectively. The resulting total SQUID junction area averages $0.056 \pm 0.004\ \mu\mathrm{m}^2$ (untrimmed) versus $0.058 \pm 0.003\ \mu\mathrm{m}^2$ (trimmed, $p = 0.057$). Because \ABAA{} is an electrical post-fabrication treatment that modifies barrier coordination without altering lithographic perimeter boundaries~\cite{Pappas2024ABAA}, these sub-$3\%$ dimensional offsets arise purely from the non-uniform allocation of test devices across design targets rather than from the annealing process itself.

\begin{figure*}[tb]
\centering
\includegraphics[width=0.9\textwidth]{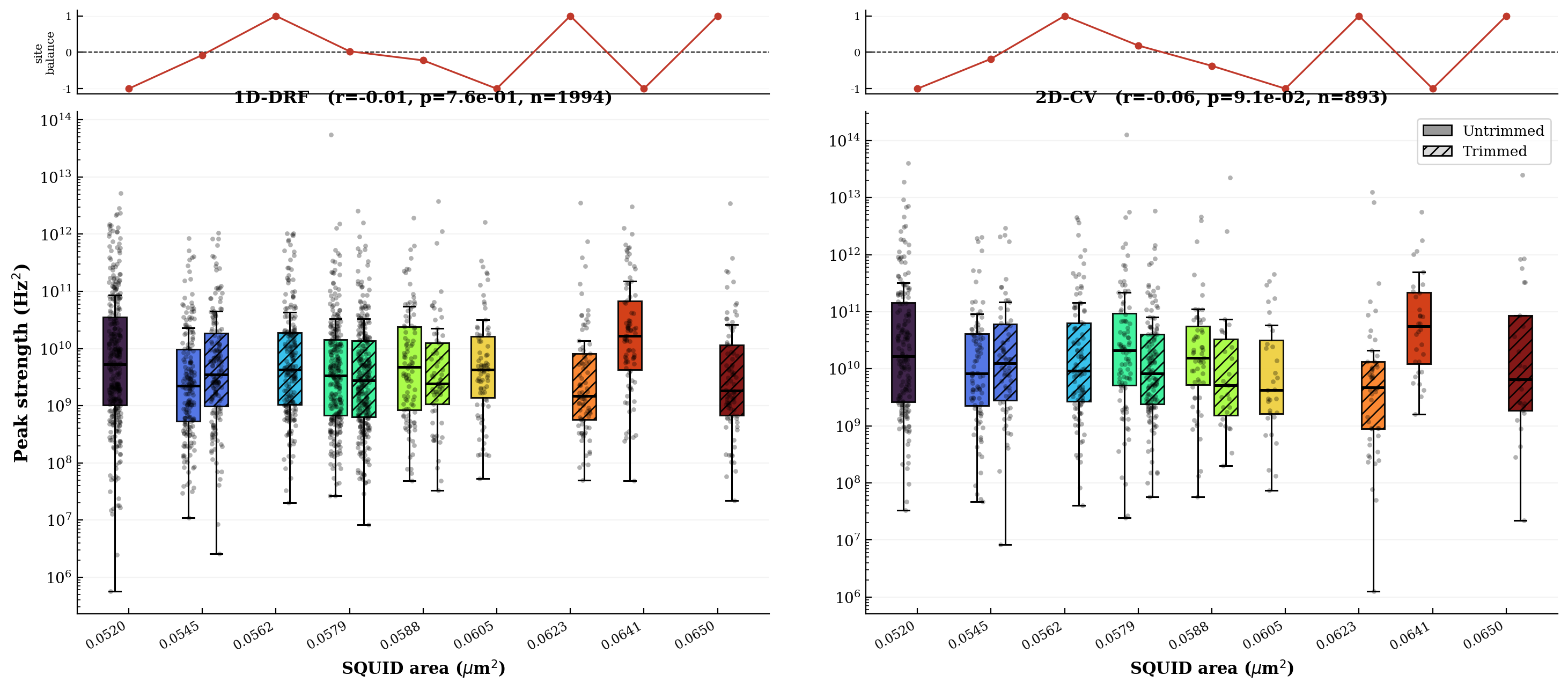}
\caption{\textbf{Lorentzian TLS loss integrals resolved across discrete SQUID junction areas.} 
Frequency-integrated loss strengths $\mathcal{I}_{\mathrm{TLS}}$ ($\mathrm{Hz}^2$) grouped across nine discrete lithographic SQUID areas ($0.0520$ to $0.0650\ \mu\mathrm{m}^2$) for \DRF{} ($n = 1994$ peaks, Pearson $r = -0.01$, $p = 0.76$, left) and \TL{} ($n = 893$ peaks, Pearson $r = -0.06$, $p = 0.091$, right). Solid and hatched boxes denote untrimmed and trimmed device subsets, respectively; top traces illustrate the relative sample balance across design targets. Defect dissipation exhibits no systematic scaling with junction area across either diagnostic pipeline.}
\label{fig:jj-box}
\end{figure*}

To test whether this geometric variation influences defect dissipation, we first stratify the fitted Lorentzian loss integrals $\mathcal{I}_{\mathrm{TLS}}$ by discrete junction area (Fig.~\ref{fig:jj-box}). Across both pipelines, defect loss distributions remain stationary across design categories, exhibiting no systematic correlation with SQUID area ($r = -0.01$ for \DRF{}, $r = -0.06$ for \TL{}).

\begin{figure*}[tb]
\centering
\includegraphics[width=1.0\textwidth]{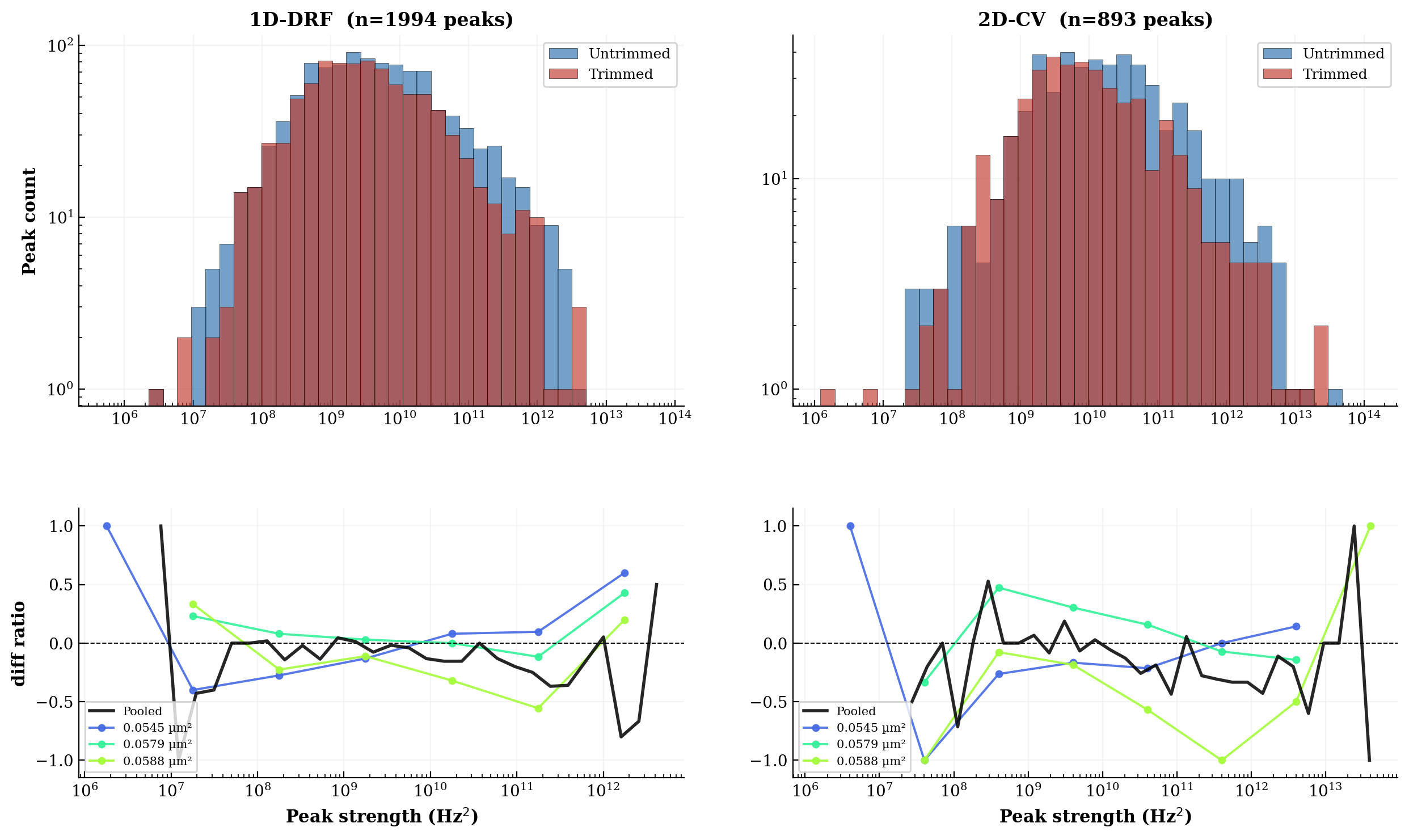}
\caption{\textbf{Area-resolved differential analysis of TLS loss suppression.} 
Top row: Pooled empirical loss-strength histograms comparing untrimmed (blue) and trimmed (red) transmon ensembles for \DRF{} ($n = 1994$ peaks, left) and \TL{} ($n = 893$ peaks, right) on logarithmic scales. 
Bottom row: Treatment differential ratio curves, defined as $(\mathcal{P}_{\mathrm{trim}} - \mathcal{P}_{\mathrm{untrim}}) / (\mathcal{P}_{\mathrm{trim}} + \mathcal{P}_{\mathrm{untrim}})$, evaluated for the pooled population (solid black curve) and within individual SQUID area designs containing balanced sample allocations ($0.0545\ \mu\mathrm{m}^2$, $0.0579\ \mu\mathrm{m}^2$, and $0.0588\ \mu\mathrm{m}^2$). The systematic depletion of high-dissipation defects ($\mathcal{I}_{\mathrm{TLS}} > 10^{11}\ \mathrm{Hz}^2$, negative differential ratio) is consistently reproduced within isolated junction geometries.}
\label{fig:jj-robustness}
\end{figure*}

We next evaluate the treatment differential within matched junction-area subsets containing both untrimmed and trimmed devices (Fig.~\ref{fig:jj-robustness}). The preferential suppression of high-loss defects ($\mathcal{I}_{\mathrm{TLS}} > 10^{11}\ \mathrm{Hz}^2$) persists when comparisons are restricted to devices sharing identical junction lithography ($0.0545$, $0.0579$, and $0.0588\ \mu\mathrm{m}^2$). This confirms that the observed loss reduction is not an artifact of sample allocation across design geometries.

Taken together, these checks confirm that the observed count--loss decoupling is an intrinsic physical consequence of the low-to-moderate $\sim10\%$ \ABAA{} treatment rather than an effect of lithographic junction dimensions.

\clearpage
\bibliographystyle{apsrev4-2}
\bibliography{tls_refs_v2}

\end{document}